\documentclass[12pt]{article}
\usepackage{epsfig}

\def\beq{\begin{equation}}
\def\eeq#1{\label{#1}\end{equation}}
\def\eeqn{\end{equation}}

\def\beqa{\begin{eqnarray}}
\def\eeqa#1{\label{#1}\end{eqnarray}}
\def\eeqan{\end{eqnarray}}

\let\bar=\overbar

\def\O{{\cal O}}

\def\Dslash{\not{\hbox{\kern-4pt $D$}}}
\def\dslash{\not{\hbox{\kern-2pt $\del$}}}

\def\ee{e^+e^-}

\def\msb{{\bar{\ssstyle M \kern -1pt S}}}

\def\Title#1{\begin{center} {\Large {\bf #1} } \end{center}}

\begin{document}
\newcommand \ga{\raisebox{-.5ex}{$\stackrel{>}{\sim}$}}
\newcommand \la{\raisebox{-.5ex}{$\stackrel{<}{\sim}$}}
\def\be{\begin{equation}}
\def\ee{\end{equation}}
\def\bea{\begin{eqnarray}}
\def\eea{\end{eqnarray}}

\Title{Neutron Star Masses, Radii and Equation of State}

\bigskip\bigskip

%+\addtocontents{toc}{{\it H. Heiselberg}}
%+\label{ReggianoStart}

\begin{raggedright}  

{\it Henning Heiselberg\index{Heiselberg, H.}\\
Nordita\\
Blegdamsvej 17\\
DK-2100 Copenhagen \O , Denmark}
\bigskip\bigskip
\end{raggedright}

\section{Introduction}

We are closing in on neutron stars both observationally and theoretically.
Observationally, a number of masses (M) and a few radii (R) have been
measured as well as a number of other properties.
Theoretically, modern equation of states (EOS) are more reliable 
due to precision measurements of nucleon-nucleon interactions,
detailed calculations of binding energies of light nuclei and nuclear
matter which constrain
three-body forces,
inclusion of relativistic effects, improved many-body and
Monte Carlo methods, etc.

Ultimately, we can exploit the one-to-one correspondence
%\footnote{subtleties as double valuedness is not a problem.}
between the EOS and the mass-radius relation of cold stellar object:
\bea
  P(\rho)\quad \Leftrightarrow \quad M(R)
\eea
Observing a range of neutron star M and R thus
reveals the EOS (e.g., pressure P versus density $\rho$) of dense 
and cold hadronic matter.
Possible phase transitions from nuclear matter to quark matter (either can
also undergo superfluid transitions at certain densities and temperatures), 
hyperon matter, kaon or pion condensates, etc.,
would also be revealed by an anomalous/kinky function $P(\rho)$ and $M(R)$.
The higher the order of the transition is, the smoother will $M(R)$ be
and very accurate observations will thus be necessary. 
On the other hand, just one accurately measured neutron star mass and radius
would already constrain the EOS significantly.
Information on the EOS at high baryon densities and the presence or absence
of phase transitions 
could guide us in solving QCD after decades of unsuccessful attempts.

In the following I shall give a brief account of the present status on
neutron star
observations and theory referring to \cite{Latt,Balb,annrev,physrep} for
longer reviews. Subsequently, I shall attempt to recount
the most likely possible phase transition in dense nuclear matter
with emphasis on quark matter and its possible color superconducting
states, as this is most relevant at this conference.
Finally, I shall point to important developments expected in the
near future. 

\begin{figure}[htb]
\begin{center}
\epsfig{file=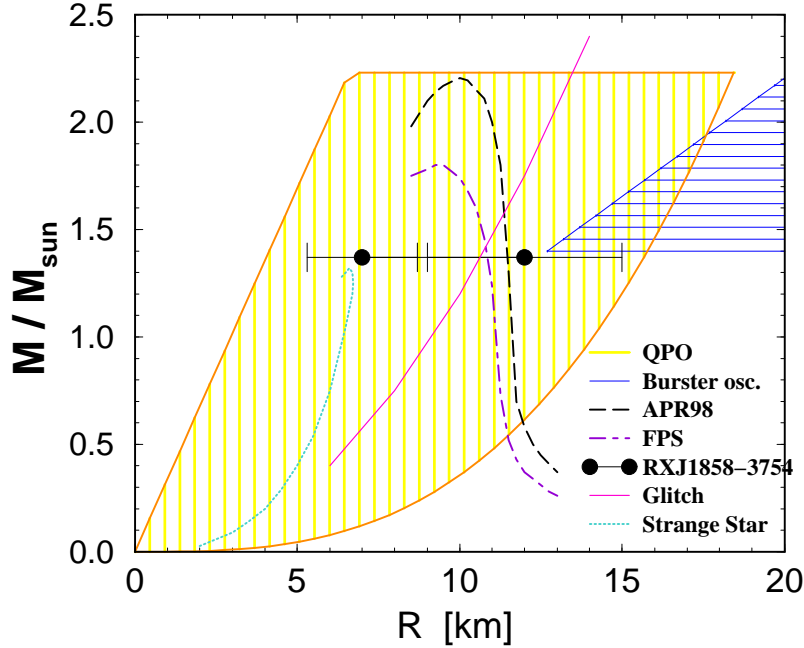,height=3.5in}
\caption{Neutron star masses vs.~radius for modern \cite{apr98}
and FPS \cite{FPS} EOS and strange stars \cite{Bombaci}.
The hatched areas represent the neutron star
radii and masses allowed for 
orbital QPO frequencies 1060~Hz of 4U 1820-30 (vertical lines, 
\cite{zss97,miller})
and for burster oscillations of 4U 1636-53 assuming
$M\ge 1.4M_\odot$ (horizontal lines, \cite{Nath}) area.
Models for glitches in the Vela pulsar constrain masses and radii
\cite{Link} below the full line.
The radii of RX J1856-3754 from Refs. \cite{Pons,Kerk} assumes
$M=1.37M_\odot$.}
\label{MR}
\end{center}
\end{figure}

\section{Observed neutron star masses}
Only a few masses have been determined from the more than thousand neutron
stars, that have been discovered so far:

{\bf Binary pulsars:}
Six double neutron star binaries are
known so far, and all of them have masses in the
surprisingly narrow range $1.36\pm 0.08 M_\odot$ \cite{Thor99}. 
Neutron stars are estimated to have a binding energy of $\sim 10 \%$
of their mass.  Thus $\sim 1.5$ $M_\odot$ of nuclei are needed to
obtain a $1.35$ $M_\odot$ star. It is suspicious that the Chandrasekhar
mass (maximum mass before gravitational collapse sets in) for the iron core
inside a large burning star is just around $\sim 1.5$ $M_\odot$. It is
therefore
a tempting conclusion that the iron cores are the progenitors of neutron
stars and that all neutron stars are simply produced with 
$M\simeq1.35$ $M_\odot$.
Similarly, white dwarfs are formed in a narrow mass range around
$M\simeq 0.6$ $M_\odot$ whereas
their Chandrasekhar mass is $M\simeq1.35$ $M_\odot$. The latter mass
is probably reached by accretion and is responsible for supernova
type SN-1a used as standard candles in cosmology. 
Calculations of supernova explosions do, however, indicate
that neutron stars should be formed with a wide range of masses
and so the narrow binary pulsar mass range could be a selection effect
in forming a double neutron star system.
Whereas  selection effects often are important in
astrophysics the contrary is the case
in particle physics, where one does not believe in accidents but invoke
some underlying symmetry.

Millisecond binary pulsars with
white dwarf companions have less precise mass determination.
Promising progress is reported, e.g., for PSR J0437-4715 \cite{vanstraten}
where a pulsar mass of $M=(1.58\pm0.18)M_\odot$ (error bars are $2\sigma$)
is found.

{\bf Vela X-1} and {\bf Cygnus X-2} are X-ray binary pulsar/burster
with high/low mass companions respectively. From X-ray pulse delays,
optical radial velocities and constraints from X-ray eclipse, their
masses have been determined.
For Vela X-1: $M=1.87^{+0.23}_{-0.17}\ M_\odot$ \cite{Barziv},
and for Cygnus X-2: $M=1.8\pm0.4M_\odot$
\cite{Orosz}.

{\bf QPO's} are neutron stars emitting X-ray's at frequencies of the
orbiting accreting matter.
Such {\em quasi-periodic oscillations} (QPO) have been found in
12 binaries of neutron stars with low mass companions.  If the QPO
originate from the innermost stable orbit \cite{zss97,miller} of the
accreting matter, their observed values imply that the accreting
neutron star has a mass $\simeq 2.3M_\odot$ in the case
of 4U 1820-30. If not, the QPO's constrain the EOS as shown in
Fig. 1.

\section{Observed neutron star radii}

The small size of neutron stars makes it very difficult to observe them
directly and measure their radius. Estimates have been obtained using
quite different methods, which has the benefit that the 
systematic errors are also different.

{\bf RX J1856.5-3754} is our nearest known neutron star. It is
non-pulsating and almost thermally radiating. It has been
studied recently with the Hubble space telescope by Walter et al. 
\cite{Walter}.  Its surface temperature is
$T\simeq 57$~eV and its distance is, from parallax measurements and
circumstantial evidence, about $d\sim61$~pc.  From the
measured flux
\be
   F = \sigma_{SB}T^4 R^2/d^2
\ee
one obtains a radius of $R_\infty=R/\sqrt{1-2GM/R}\simeq7$~km 
which is incompatible with almost any EOS.
Kaplan et al. \cite{Kaplan} have reanalysed the HST data and find
only half the parallax and thus twice the distance and radius
$R_\infty\simeq 15$~km corresponding to $R\simeq 12$~km for
$M=1.4M_\odot$.
Its age would be almost a million years which is compatible with standard
modified URCA cooling. The spectrum is, however, suppressed in the optical
part as compared to the X-rays.
Recent more detailed analyses of the spectrum
\cite{Pons,Burwitz} attempting to model the neutron star atmosphere
and its absorption, as well as including magnetic fields
does not improve the spectral fit.
A much better description is obtained from a two temperature model,
i.e., a small hot spot and a colder area with a larger radius
$\sim9 (d/61pc)$~km. 

Gravitational lensing by our nearest neutron star may be observed in
june 2003, when J1856.5-3754 will pass within $\sim0.3$ arcseconds of
a background star as it flies through space with proper motion of 0.332
arcseconds per year. According to \cite{Pacz} (see, however,
\cite{Kaplan}) it should just be possible to measure that the 26.5
magnitude background star moves by 0.6 milliarcseconds due to the
gravitational field of the neutron star. This accuracy
requires that the Hubble space telescope is
extended with the Advanced Camera for Surveys.  If possible, its mass
could be derived and our nearest known neutron star would then also be
the first one with both mass and radius determined.

{\bf X-ray bursts} are thermonuclear explosions of accreted matter on
the surface of neutron stars. After accumulating hydrogen on the
surface for hours, pressure and temperature become sufficient to
trigger a runaway thermonuclear explosion seen as an X-ray burst that
lasts a few seconds \cite{BildStroh}.
Assuming that the burst spectra are black-body one can from the resulting
temperature and measured flux estimate the
neutron star radius if its distance is known. 
Often the radius is underestimated because only a hot spot emits or
the spectrum contains a hard tail. Some bursts do, however, give
radii of order $\sim12$~km with a period of 
almost constant (Eddington) luminosities.

{\bf Quescent neutron stars} are non-accreting X-ray binaries of which
some are emitting thermally.
Recently, a few quescent neutron stars have been discovered
in galactic globular clusters where the distance is known 
relatively accurately. CXOU 132619.7-472910.8 in NGC 5139
has $T=66\pm5$~eV and $R_\infty=14.3\pm2.1$~km (90\% confidence limit)
and similar radii although with larger uncertainties
are obtained from half a dozen other quescent neutron stars \cite{Bildsten}.

{\bf Burst oscillations} in the X-ray flux during the first seconds of
the bursts have recently been exploited to determine the compactness
or redshift $M/R$.  Whereas the average amplitude increases with the
growing size of the hot spot, the amplitude is strongly modulated by
the rotational period of the neutron star. The flux does not
completely disappear when the hot spot is on the back side due to
bending of the light in the gravitational field. This way $M/R<0.16$
can be extracted \cite{Nath} for 4U 1636-53 and a radius
$R>12-13$~km is obtained for a $M=1.4M_\odot$ neutron star (see Fig. 1).
Corrections from aberration, doppler shifts, etc. are being
investigated.  Yet, the oscillation analyses is another new promising
method by which we can obtain neutron star masses and radii.

{\bf Absorption lines} in the neutron star photospheres should be
detectable with the spectrographs on board Chandra and XMM.  The
gravitational redshift and pressure broadening of absorption lines
determine $M/R$ and $M/R^2$ respectively and would thus complement
other mass and radius information.  First results were, however,
disappointing. Only in RX J1856.5-3754 were there indications for one
or two lines \cite{Kerk} but higher resolution is required and will come
with future observation time.

\section{Modern equation of states of dense matter}

The relation between $M(R)$ and $P(\rho)$ is often presented
(in particular by observers)
by a wide variety of curves
calculated from ``old'' EOS thus implying that one can get anything from
theory. This is very unfair as many old EOS's are inconsistent with known
nucleon-nucleon interactions and/or saturation density and binding energy
of nuclei/nuclear matter and should therefore not be taken seriously. 

Modern microscopic EOS's are actually converging \cite{physrep,annrev}. 
The NN interaction is now
well determined and constrain potential models leaving only minor differences.
Relativistic effects have been included and
current scattering experiments at intermediate energies will determine
the NN interactions at higher momenta relevant for higher densities.
Three-body interactions can be constrained in order to fit nuclear 
binding and saturation density as well as binding energies of light nuclei
up to $A\le8$.
Many-body and Monte Carlo techniques are now much more accurate.
The resulting ``modern'' EOS \cite{apr98} are quite reliable up to a few
times nuclear saturation densities. Above they are expected to break down
but can be constrained by causality conditions.
It is boldly predicted \cite{apr98} that neutron stars 
in the mass range $M\sim 0.8-1.8M_\odot$ all have radius
just around $R\simeq 11.5$~km (see Fig. 1).

The uncertainties in the EOS at densities $\rho\ga 3\rho_0$ affect
$M(R)$ for the heavy neutron stars only. By making
the EOS stiffer at high densities in a smooth way,
the maximum mass can increase up to $M\la 2.0-2.3M_\odot$ \cite{physrep}
but not much higher due to the causality ($c_s<c$) condition. 
Rotation can increase it by $\sim10$\% for the millisecond pulsars.

Phase transitions generally soften the EOS and lower the maximum mass.
If the $M=2.25M_\odot$ mass in 4U 1820-30 stands this would rule out
any major phase transition and allow only the stiffest EOS.

Many other phase transitions in neutron stars have been considered.
It is expected that
superfluid phases of neutrons and protons exist at least in certain 
density regions at low temperatures although $T_c$ have not been calculated
reliably for strongly interacting and correlated nuclear matter.
At typical neutron star densities neutrons and protons are
superfluid as well due to $^1S_0$ and, in the case of protons, also
$^3P_2$ pairing \cite{physrep}.  These superfluid and superconducting
components will have drastically different transport properties than
normal Fermi liquids.  Generally the resistance,
specific heat, viscosities, cooling, etc. are suppressed by factors of
order $\sim\exp(-\Delta_i/T)$, where $\Delta_i$ is the gap of quarks,
nucleons or electrons. A superfluid neutron gas in the inner crust is
assumed in the description of glitches \cite{Link} and provides constraints
on the EOS (see Fig. 1).

Eventually at very high densities nucleon degrees of freedom must be 
replaced by quark ones but it is not known whether core densities of
neutron stars are sufficient. Such quark stars will be discussed in the
following section.

More speculative phases are $\pi^0$, $\pi^-$ and $K^-$ condensates
as well as hyperons ($\Sigma^-$, $\Lambda$,...). In \cite{apr98} a
condensate of virtual $\pi^0$ is found in a narrow density interval
due to strong tensor correlations. The $K^-$ energy can be calculated 
at low densities and a naive extrapolation would lead to condensate
at high densities $\rho\ga4\rho_0$. However, 
correlations in nuclear matter invalidates such an extrapolation and makes
a $K^-$ condensate unlikely \cite{kaon}.
Hyperons are found to appear at rather low densities $\rho\ga 2-3\rho_0$
in a number of models \cite{Schaffner}. Due to limited information on
hyperon-nucleon two- and three-body interactions one cannot exclude the
presence of hyperons in cores of neutron stars but their effect on the
binding energy and thus the EOS is minor whereas their effect on $\mu_e$
could be substantial \cite{annrev}. 

\begin{figure}[htb]
\begin{center}
\epsfig{file=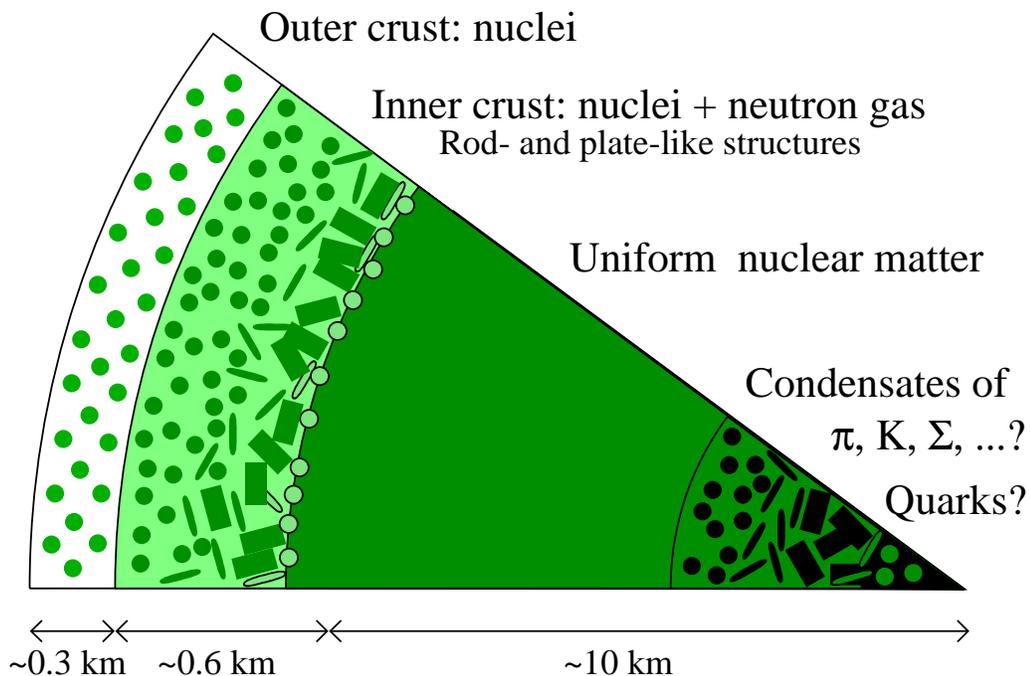,height=3.5in}
\caption{Cross section of a $\sim1.4M_\odot$ neutron star. 
The $\sim1$~km thick crust
consist of neutron rich nuclei in a lattice and a uniform background of
electrons and, in the inner crust, also a neutron gas. The interior of 
the neutron star contains a nuclear liquid of mainly neutrons and
$\sim 10$\% protons at densities above nuclear matter density
$n_0$ increasing towards the center. Here pressures and densities
may be sufficiently high that the dense cold strongly interacting
matter undergoes phase transitions to, e.g., quark or hyperon
matter or pion or kaon condensates appear.
Typical sizes of the nuclear
and quark matter structures are $\sim10^{-14}$ m but have been scaled up
to be seen.}
\label{pizza}
\end{center}
\end{figure}

\section{Quark stars}

Natures marvelous variety of EOS's results in the existence of
several different stable stellar objects all around  one solar mass. 
These can be ordinary stars, white dwarfs, neutron stars, black holes - 
and possibly also quark stars.
Quark stars come in several categories depending on the details of the
nuclear to quark matter phase transition:

\begin{itemize}
\item The pure quark stars also called
``strange'' stars consist of up, down and strange quarks with electrons
to fulfil charge neutrality. Possibly a crust of nuclei is suspended
above the surface of the quark star.
In simple bag model EOS a rather low bag constant and strange quark mass
is required to make strange stars and strangelets.
If the SAX J1808.4-3658 really has $R\simeq6$~km 
based on accretion in magnetic fields \cite{Bombaci}
or RX J1856-1754 has $R\simeq7$~km from the one-temperature fit in
Ref. \cite{Pons}
that would indicate strange stars rather than normal neutron stars.

\item Hybrid stars have a core of quark matter and a mantle of
nuclear matter. The quark core size depends on the EOS and vanishes 
for large bag constants leading to a normal (nucleons only) neutron star.

\item Mixed stars have a mixed phase of nuclear and quark matter over 
a range of density or radius. The mixed phase appear in
two-component systems, where the two components:
neutrons and protons or up and down quarks \cite{Weber}.
It is, however, required that the interface tension is sufficiently small
so that the surface
and Coulomb energies of the associated structures are small
\cite{HPS}. If not, a hybrid star results.
\end{itemize}

{\bf A mixed phase} of quark and nuclear matter has lower energy per
baryon at a wide range of densities \cite{Weber}
if the Coulomb and surface energies
associated with the structures are sufficiently small \cite{HPS,physrep}.
The mixed phase will then consist of two coexisting phases of nuclear
and quark matter in droplet, rod- or plate-like structures (see Fig. 2) in
a continuous background of electrons much like the mixed phase of
nuclear matter and a neutron gas in the inner crust of neutron
stars \cite{Lorenz}. Another requirement for a mixed phase is that the
length scales of such structures must be shorter than typical screening
lengths of electrons, protons and quarks.

In the mixed phase the nuclear and quark matter will be positively and
negatively charged respectively. $\beta$-equilibrium determines the
chemical potentials and densities in the coexisting phases.
Total charge neutrality 
\be
  n_e=(1-f)n_p+f(\frac{2}{3}n_u-\frac{1}{3}n_d-\frac{1}{3}n_s) \,, \label{mix}
\ee
where $n_p$ is the proton density, determines 
the ``filling fraction'' $f$,
i.e. the fraction of the volume filled by quark matter. For pure
nuclear matter, $f=0$, the nuclear symmetry energy can force the
electron chemical potential above $\sim100$ MeV at a few times normal
nuclear matter densities. With increasing filling fraction, however,
negative charged droplets of quark matter replace some of the
electrons and $\mu_e$ decreases. With increasing density and filling
fraction it drops to its minimum value given $\mu_e=m_s^2/4\mu$
corresponding to pure quark matter, $f=1$.

\begin{figure}[htb]
\begin{center}
\epsfig{file=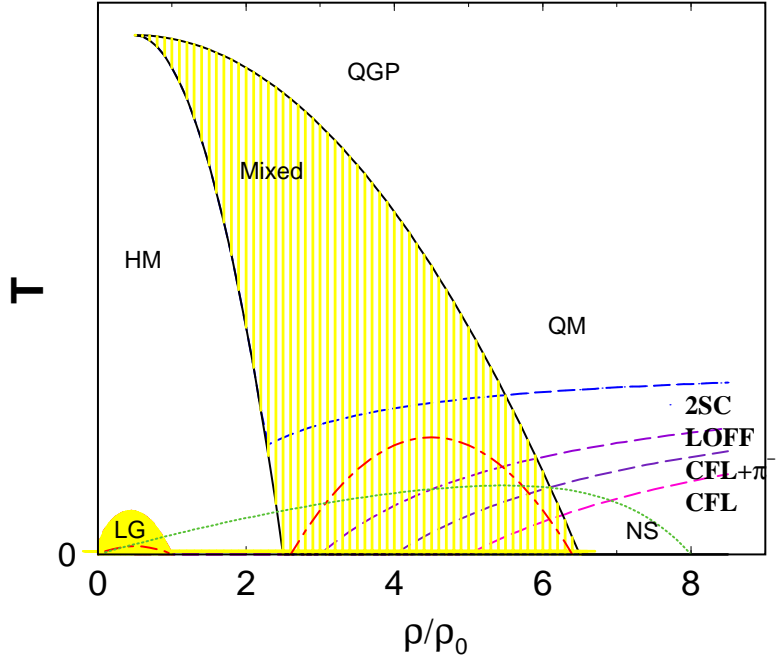,height=3.5in}
\caption{Sketch of the QCD phase diagram, temperature vs. baryon density 
{\em in neutron star matter}, i.e. charge neutral and in $\beta$-equilibrium
containing electrons. Hatched areas indicate 
mixed phases of hadronic matter (HM) and
quark matter (QM/QGP) as well as the nuclear liquid-gas. 
Dash-dotted lines indicate melting
temperatures of the lattices in the mixed phase.
Dashed lines separate CSC phases that may appear (see text). 
The trajectory of neutron
star core densities during formation is shown by dotted line and densities
below it exist inside neutron stars.}
\label{phase}
\end{center}
\end{figure}

\section{Color superconductivity}
If quark matter exists in neutron stars or is produced in heavy-ion
collisions, a condensate of quark
Cooper pairs may appear at low temperatures
characterized by a BCS gap $\Delta$
usually referred to as color
superconductivity (CSC) \cite{Alford}. 
The appearance of a gap through color-flavor
locking (CFL) requires the gap to exceed the difference between the
{\em u,d,s} quark Fermi momenta, which is not the case for sufficiently
large strange quark masses or for an
appreciable electron chemical potential, $\mu_e$, which is present in
neutron star matter as discussed in \cite{HCSC}.

In neutron star matter $\beta$-equilibrium relates the quark
and electron chemical potentials
\be
   \mu_d=\mu_s=\mu_u+\mu_e \,. \label{chem}
\ee
Temperatures are normally much smaller than typical Fermi energies in
neutron stars. 

The strange quark masses and electron chemical potentials stress the system
in the direction of splitting the quark chemical potentials. The pairing
interaction prefer overlapping quark Fermi surfaces. We shall investigate
this competition in detail below. The case where 
interactions between quarks are strong and the effective 
pairing gaps is larger than $\Delta\ga \sqrt{2}|\mu_e/2-m_s^2/8\mu|$
was shown in \cite{Rajagopal} to favor the CFL phase.

If interactions are weak, 
the chemical potentials are then related to Fermi momenta
by $\mu_i=\sqrt{m_i^2+p_i^2}$. 
If the strange quark mass $m_s$ is  much smaller than the 
quark chemical potentials, Eq. (\ref{chem}) implies
a difference between
the quark Fermi momenta
\begin{eqnarray}
   p_u-p_d &=& \mu_e                           \,, \label{ud} \\
   p_u-p_s &\simeq& \frac{m_s^2}{2\mu} -\mu_e  \,, \label{us} \\
   p_d-p_s &\simeq& \frac{m_s^2}{2\mu}         \,, \label{ds} 
\end{eqnarray}
where $\mu$ is an average quark chemical potential.
Strange quark masses are estimated from low energy QCD
$m_s\simeq150-200$~MeV and typical quark chemical potentials are
typically $\mu\simeq 400-600$~MeV for quark matter in neutron
stars \cite{physrep}. Consequently,
$m_s^2/2\mu\simeq 10-25$~MeV.

The BCS gap equation has previously been solved for pure
{\em u,d} and
{\em u,d,s} quark matter ignoring electrons and $\beta$-equilibrium 
and the conditions for condensates of dicolor pairs (2CS) and CFL
respectively were obtained. 
%\cite{Love,Wilczek,Shuryak,ABR}.
There are three 2CS conditions
\be 
  \Delta \ga |p_i-p_j| \,,\quad i\ne j=u,d,s \label{CFL} \,. 
\ee
In bulk quark matter total charge neutrality of quarks and electrons
and $\beta$-equilibrium require that $\mu_e \simeq m_s^2/4\mu$.
In a mixed phase of quark and nuclear matter the electron
chemical potential is a decreasing function from the value in
pure $\beta$-equilibrium nuclear matter $\mu_e\sim100-200$~MeV
down to that for bulk quark matter (if the cores of neutron
stars are very dense) $\mu_e \simeq m_s^2/4\mu \simeq 5-10$~MeV.

We can now give a qualitative picture of the various phases that
strongly interacting matter undergoes as the density increases towards
the center of a cold neutron star (see Fig. 3). The inner crust will undergo
several transitions as the nuclear matter and neutron gas mixed phase
change dimensionality via nuclei, rods, plates, tubes, bubbles to
nuclear matter in which the neutron and protons may be superfluid.  If
quark matter appears at higher densities in a mixed phase with nuclear
matter those structures repeat again.

Inside the quark matter part of this mixed phase the CSC phase also
changes with density or $\mu_e$ depending on the size of the effective
pairing interaction $\Delta$
between two quarks {\it i,j} (see, e.g., \cite{Alford,diq}). 

\begin{itemize}

\item
If the pairing is strong $\Delta\ga \sqrt{2}|\mu_e/2-m_s^2/8\mu|$
the CFL phase is favored and no electrons appear as
was shown in \cite{Rajagopal}. This condition is not likely to be fulfilled
in the mixed phase with low quark filling fraction, where $\mu_e$ may be of
order $100-200$~MeV or larger.

\item
If $\mu_e/2\ga\Delta\ga m_s^2/2\mu$ a number of
CSC phases appear. In the beginning of the mixed phase ($f\sim0$) only
(\ref{ds}) fulfils (\ref{CFL}) and we have a 2CS of {\em d,s}
quarks. A 2CS of {\em u,s} quarks may, however, compete at larger
filling. At the end of the mixed phase ($f\sim1$) all
(\ref{ud},\ref{us},\ref{ds}) fulfil (\ref{CFL}) resulting in CFL.
In between a crystalline LOFF phase and a CFL phase with a $\pi^-$
condensate (analogous to the CFL-$K^0$ phase in \cite{Alford}) may
appear. 

\item
If the pairing interaction is weak or the strange quark mass large
such that $\Delta\la m_s^2/4\mu$ a number of different CSC phases
such as a 2CS of {\em u,s} quarks or 
CFL-K may appear as has been discussed in \cite{Alford,Splittorp}.

\end{itemize}

In the mixed phase gaps may be affected by the finite size of the
quark matter structures as is the case for nuclei. Also surface
and Coulomb energies generally disfavors mixed phases.

The finite temperature extension of the competing phases, calculations
of the critical temperatures and densities, order of the
transitions, etc. for neutron star matter should be investigated further.
Probably, the superfluid phases undergo a second order phase transition
to the normal phase with increasing temperature at constant baryon density.
However, transitions between competing phases might also occur as indicated in
Fig. 3.

Temperatures in neutron
stars, $T\la 10^6K\simeq 10^{-4}$~MeV after cooling, 
are typically much lower than lattice 
melting temperatures, $T_{melt}\simeq Z^2e^2/170a$ where $a$ is the
lattice spacing (see Fig. 3). Thus the quark matter
structures would be solid frozen and the cores of neutron stars
would be crystalline and possibly also CSC.
Lattice vibration
will couple electrons at the Fermi surface with opposite momenta and
spins via phonons and lead to a ``standard'' BCS gap for electrons.
The isotopic masses are similar but as densities and Debye frequencies
are larger, we can expect considerably larger BCS gaps for
electrons. The electrical superconductivy affect magnetic fields
through the Meissner effect and magnetic field decay.

\section{Outlook}
A number of promising developments have been mentioned above 
that may provide new information
on neutron star masses, radii, EOS and possible phase transitions in
high density matter.

On the observational front more and better masses and radii are
determined by a number of different methods (thus with different
systematic errors). On the theoretical front the uncertainties in the EOS
are reduced by improved two- and three-body forces, relativistic effects
and manybody calculations leading to socalled modern EOS.
A number of
phase transitions are still possible at high densities
and can lead to marvelous structured phases and condensates. 

We can hope for additional information from other directions as well:

{\bf Gamma ray bursters}.
The discovery of afterglow in {\em Gamma Ray Bursters} (GRB)
allows determination of their very high redshifts ($z\ge 1$).  They
imply that GRB have an enormous energy output $\sim
10^{53}$ ergs which requires some central engine more powerful
than ordinary supernovae \cite{GRB}.
These could be a special class of type Ic supernova ({\it hypernovae})
where cores collapse to black holes, or binary neutron stars merging,
or some major phase transition to, e.g., quark matter \cite{Sannino}.
Also soft GRB may be explained by accretion on strange stars \cite{Usov}.

{\bf Neutrinos} from the formation of a ``proto-neutron star''
will be detected from the predicted 1-3 supernovae explosions in our
and neighboring galaxy per century.  In the case of the recent 1987A
in LMC 19 neutrinos were detected on earth and with the
upgraded neutrino detectors many thousand neutrinos are expected.  The
neutrinos can test the SN models, the neutron star EOS and early
cooling. 
During the first second rapid cooling takes place and a 
phase transitions to, e.g., quark matter or CSC may
occur. This may result in a delayed neutrino
blimps \cite{Greg}.

{\bf Gravitational wave} ground based interferometic
detectors currently under commission will improve
sensitivity significantly. Detectable candidate sources are inspiralling
binary neutron stars or black holes merging which may be responsible for
gamma ray bursts. R-mode instabilities in rapidly rotating neutron
stars may also be detectable \cite{Madsen,LIGO}.

{\bf Relativistic heavy-ion collisions} are probing the high
temperature and low baryon density part of the QCD phase diagram at
the opposite end to neutron stars. The phase transition from hadronic
matter to a quark-gluon plasma is searched for though the transition
may well be a smooth cross over. Some ``anomalous'' effects, i.e.
deviations from predictions based on hadronic theory, have been
found and are currently studied intensively at RHIC. It has been
speculated that one might find a critical point at high temperatures
which would then indicate that the smooth cross over changes to a
first order transition at higher baryon densities. That would be
valuable information for neutron stars although it would not tell at
which density the transition would occur at low temperatures.

{\bf Superfluidity in cold fermionic atomic systems}
will, due to the rapid progress
in cooling and trapping techniques, soon be discovered.
The great advantage of atomic traps is that one can vary the number of
particles, their density and interaction strength as well as the
number of spin states, masses, etc.
 Measuring the size of the pairing gap and varying these parameters
provides a testing ground for analytical
calculations in the dilute or weakly interacting limit \cite{gap}.
This would be very useful for gap calculations in neutron and nuclear
matter as well as quark matter.

{\em The future} of neutron star observations looks bright as new windows are
about to open. A new fleet of X- and Gamma-ray satellites have and
will be launched. With upgraded ground based observatories and
detectors for neutrinos and gravitational waves \cite{LIGO}  
our knowledge of neutron star properties will be greatly improved.
Heavy-ion physics at RHIC and LHC may add further to our understanding
of the QCD phase diagram. 

\bigskip
Thanks to R. Ouyed and F. Sannino for organizing this 
conference and comments on the manuscript.

\def\Discussion{
\setlength{\parskip}{0.3cm}\setlength{\parindent}{0.0cm}
     \bigskip\bigskip      {\Large {\bf Discussion}} \bigskip}
\def\speaker#1{{\bf #1:}\ }
\def\endDiscussion{}
 
\end{document}